\documentclass[aps,superscriptaddress,showpacs,amsmath,amssymb,amsfonts,altaffillsymbol,twocolumn,notitlepage,prl]{revtex4-2}

\usepackage{wasysym}

\usepackage{graphicx}
\usepackage{dcolumn}
\usepackage{bm}
\usepackage{ulem}
\usepackage{amsmath}
\usepackage{color,xcolor}

\begin{document}

\preprint{APS/123-QED}

\title{Discontinuous rigidity transition associated with shear jamming in granular simulations}

\author{Varghese Babu}
\email{varghese@jncasr.ac.in}
\affiliation{
Jawaharlal Nehru Center for Advanced Scientific Research, Jakkur Campus, Bengaluru 560064, India.
}
\author{H. A. Vinutha}
\affiliation{Department of Physics, Institute for Soft Matter Synthesis and Metrology, Georgetown University, Washington, DC, USA.}
 \author{Dapeng Bi}
 \affiliation{Department of Physics, Northeastern University, MA 02115, USA}
\author{Srikanth Sastry }
\email{sastry@jncasr.ac.in}
\affiliation{
Jawaharlal Nehru Center for Advanced Scientific Research, Jakkur Campus, Bengaluru 560064, India.
}
\date{\today}

\begin{abstract}


We investigate the rigidity transition associated with shear jamming in frictionless, as well as 
frictional, disk packings in the quasi-static regime and at low shear rates. For frictionless disks, the transition is under quasistatic shear is discontinuous, with an instantaneous emergence of a system spanning rigid cluster at the jamming
transition. For frictional systems, the transition appears continuous for finite shear rates, but becomes sharper for lower shear rates. In the quasi-static limit, it is discontinuous as in the frictionless case.  Thus, our results show that the rigidity transition associated with shear jamming is discontinuous, as demonstrated in a past for isotropic jamming of  frictionless particles, and therefore a unifying feature of the jamming transition in general. 


\end{abstract}

\maketitle


Granular materials can exist in a flowing or a solid state. The transition between 
these states, called the jamming transition, has been the subject of intense research 
\cite{behringer2018physics,zhang2005jamming,van2009jamming}, particularly  
under isotropic compression of frictionless sphere packings.
The jamming point $\phi_J$ for packings of soft particles exhibits many characteristics of a second-order phase transition, at which various quantities show power law scaling -- with respect to the distance from the jamming point --  as one compresses beyond the jamming point \cite{ohern2003jamming,charbonneau2015jamming}. 
Further, the distribution of small forces between particles  just in contact, as well as the gaps between particles nearly in contact, also exhibit power law behavior. Exponents characterizing these are constrained by an inequality that is saturated for configurations at the jamming point, which are therefore ``marginally stable'' \cite{wyart2012marginal,lerner2013low}. The 
mean-field theory of glasses and jamming has predictions for these exponents which 
match numerical values for dimensions $D=2$ and above\cite{charbonneau2015jamming}. Extensions of this theory predict these exponents to be the same for shear jamming, as recent numerical results indeed confirm, along with the aforementioned aspects of criticality \cite{babu2022criticality}. These and related results \cite{babu2022criticality,baityjesi2017emergent} strongly support a unified description of  both isotropic and shear jamming. 



In contrast, the manner in which the contact network acquires rigidity is strongly discontinuous 
\cite{ellenbroek2015rigidity,ortiz2022assur} for frictionless isotropic jamming. At the jamming point, the entire system (barring a small 
percentage of {\it rattlers}, described later) acquires rigidity discontinuously. From the Maxwell criterion for the 
rigidity of networks of nodes connected by edges representing  distance constraints, the contact network of a configuration 
with $N$ particles in $D$ dimensions can be rigid when contacts result in at least $N_c = D(N-1)$ constraints on the non-
global degrees of freedom. In general, this is a necessary but not sufficient condition. Therefore, isotropic 
jamming occurs at the {\it isostatic point}, where the system has just the minimum number of contacts per particle, $Z$ required, 
$Z_{iso}=2D$ (from ${N Z_{iso} \over 2} = N D)$). This discontinuous rigidity transition is different 
from the  continuous transition observed, {\it e. g.} for {\it sticky} packings \cite{tighe2018,tighe2020}, and 
in random spring networks  {\cite{jacobs1996generic,chubynsky2006self}} for which the rigid component of the system grows continuously beyond rigidity percolation, which does not occur at the 
isostatic point, and is preceded by the presence of both rigid and over-constrained regions.




Results available for shear jamming appear to suggest that the rigidity transition is continuous, in contrast to isotropic jamming\cite{henkes2016rigid,liu2019frictional,liu2021spongelike,vinutha2019force}. Computational investigation of  the rigidity transition for frictional two dimensional ($2D$) systems sheared at finite rates  \cite{henkes2016rigid} revealed a broad distribution of rigid cluster sizes with increasing mean size as the jamming transition is approached, supporting a continuous rigidity transition, although becoming ``sharper'' as the shear rate is lowered. Similar results have been recently reported from analysis of sheared granular packings in experiments \cite{liu2021spongelike}. Following the observation that sheared {\it frictionless packings} acquire geometric characteristics associated with jamming \cite{vinutha2016disentangling}, the rigidity transition in such packings in $2D$ was analysed by including constraints associated with friction \cite{vinutha2019force}. The size distribution of overconstrained clusters, similar to \cite{henkes2016rigid}, exhibits a broad distribution, supporting a continuous rigidity transition. In addition, the rigidity transition associated with jamming in  frictional systems were studied in lattice models of jamming where a continuous transition was observed except in a limiting case corresponding to infinite friction \cite{liu2019frictional}.

These observations suggest that the nature of the rigidity transition could be an exception to the commonality of isotropic and shear jamming phenomenology outlined earlier. In this letter, we therefore investigate carefully the nature of the rigidity transition for both sheared frictional and frictionless packings, under both quasi-static and at finite shear rate. We find that the rigidity transition is unambiguously discontinuous under quasistatic shear. Such a transition appears rounded in the case of finite shear rate, but the dependence on shear rate clearly supports an approach to a discontinuous transition in the limit of vanishing shear rate. 

Shear jammed frictionless packings  are obtained by shearing unjammed bidisperse  soft-disk  mixtures of size ratio $1:1.4$ above the minimum jamming density $\phi_J$. As described in \cite{babu2021dilatancy,babu2022criticality,das2020unified,kumar2016memory}, well annealed disk-packings jam at a packing fraction higher than $\phi_J$. In this study, we equilibrate hard-disk configurations at high density, which jam at a density $\phi_j>\phi_J$ with the protocol described in \cite{ohern2003jamming}. Unjammed configurations decompressed to a density $\phi$, with  $\phi_J<\phi<\phi_j$ undergo shear jamming when subjected to athermal quasistatic shear (AQS) wherein strain increments of $\Delta \gamma = 5\times 10^{-4}$ are applied, each step followed by energy minimization. We study $3$ independent samples of $N=16384$ particles at a density of $\phi=0.8485$, with $\phi_j \approx0.85$ and $\phi_J\approx0.84$. 

We use Discrete Element Method (DEM) \cite{cundall1979discrete} to simulate frictional disks, using LAMMPS \cite{plimpton1995fast}, with linear and tangential spring dash-pot forces. The model includes damping in both normal and tangential directions, in addition to global viscous damping. The normal and tangential spring constants $k_n$ and $k_t$ are set to $2.0$. The normal velocity damping $\eta_n$ is set to $3.0$ and the tangential damping $\eta_t$ is set to $\frac{1}{2}\eta_n$. The global damping term $\eta$ is also set to $\approx 3$.  

Shear is applied by performing an affine transformation of particle positions, with strain increments   $\Delta \gamma$ followed by relaxation using DEM. Because of the damping terms, the system will eventually reach a force, torque balanced configuration if one waits long enough. Quasistatic shear requires reaching force/torque balance at each strain step. 
In practice,  we consider the system to have reached force/torque balance when the total force (sum of total forces 
 acting on the disks) is less than $10^{-11}$ or when the total kinetic energy of the 
 system is less than $10^{-19}$.  The simulation is stopped when the number of timesteps 
 reaches $2\times10^{9}$ regardless. The timescale required to relax the system diverges at the shear jamming transition as pointed out in \cite{vinutha2020timescale} and thus it is difficult to achieve force-balance close to the transition. For a finite shear rate $\dot{\gamma}$, each strain step is followed by DEM dynamics of duration $\Delta {\gamma}/\dot{\gamma}$.
We set $\Delta \gamma$ is $10^{-4}$ for finite rate shear and $10^{-3}$ for quasi-static shear.
We perform finite rate shear on a system size of $N=16384$ particles for $10$ independent samples (and $20$ samples for highest and lowest shear rate), and quasi-static shear with $N=2000$ for $16$ samples. The packing fraction 
$\phi$ of the system is $0.81$. 
Further details of the simulations are described in the supplemental material (SM) section I 
\footnote{See Supplemental Material at https://xxx for the additional information on:  
    I) Details of frictional shear simulation, 
    II) Pebble game algorithm, 
    III) Identifying rattlers,
    IV) Percolation analysis of the rigid clusters, 
    V)  Results with smaller friction coefficient $\mu=0.1$,
    VI) Identification of mobilized contacts,
    VII) Over-constrained networks and floppy modes.}.  
    We describe the results for friction coefficient $\mu=1$ in the main text. Results with $\mu=0.1$ can be found in SM section V.


A major distinction between frictionless and frictional jamming is the isostatic contact number $Z$ at which jamming can occur in the absence of redundant constraints, 
which has been shown to range from $D+1$ to $2D$ depending on the friction co-efficient $\mu$ 
\cite{shundyak2007force,henkes2010critical,vinutha2016disentangling,vinutha2019force}  with $Z_{iso} = D+1$ for  $\mu=\infty$. 
This can be understood using the generalized 
isostaticity condition, obtained by considering additional   
conditions due to the ``mobilized contacts''\cite{shundyak2007force}.  The tangential frictional force between two particles has an upper 
bound due to the Coulomb threshold: $f_t \leq \mu f_n$ and the mobilized contacts are those  for which $\frac{f_t}{f_n} \approx \mu$. Considering a configuration with $N$ particles and $n_{m}N$ mobilized contacts, the  conditions that the contact network at jamming has to satisfy  are $DN$ force balance 
conditions, $\frac{D(D-1)}{2}N$ torque balance conditions and $n_mN$ Coulomb conditions. The 
number of constraints imposed by the contacts is ${N D Z \over 2}$ (since each contact constrains one translational and $D-1$ rotational degrees of freedom). $Z$ is by default computed excluding {\it rattlers} (particles with less than the minimum number of contacts necessary for local rigidity, $=3$ for frictionless, and $2$ for frictional particles in 2D), and represented by $Z_{NR}$ for clarity. Defining $Z_{\mu} = Z_{NR} - \frac{2 n_m}{D}$, the generalized iso-staticity condition is 
\begin{equation}
        Z_{\mu}^{iso} = Z_{NR} - \frac{2 n_m}{D} = D+1.
\end{equation}

For $2D$ networks arising in several contexts including jamming, the onset of rigidity has been analysed by employing the pebble game algorithm\cite{jacobs1996generic}. Each node of the network represents a disk in the present context and is assigned $k$ pebbles ($k = 2$ for frictionless disks and $k = 3$ for frictional disks) representing the degrees of freedom. The constraints imposed by each contact are represented by $1$ or $2$ edges ($2$ for the frictional case, $1$ for the frictionless case, as well as for a mobilised contact). A $(k,l)$ pebble game ($l =2$ indicates the global degrees of freedom) assigns pebbles recursively to edges, and based on such an assignment, decomposes the network into rigid clusters that are mutually floppy. Rigid clusters with redundant bonds (with no assigned pebbles) are termed over-constrained. A more detailed description of the algorithm is provided in SM section II. We employ the pebble game to monitor the size of the largest rigid cluster in the system primarily, as well as the distribution of the size of rigid clusters. 

\begin{figure}
    \centering
    \includegraphics[scale=0.65]{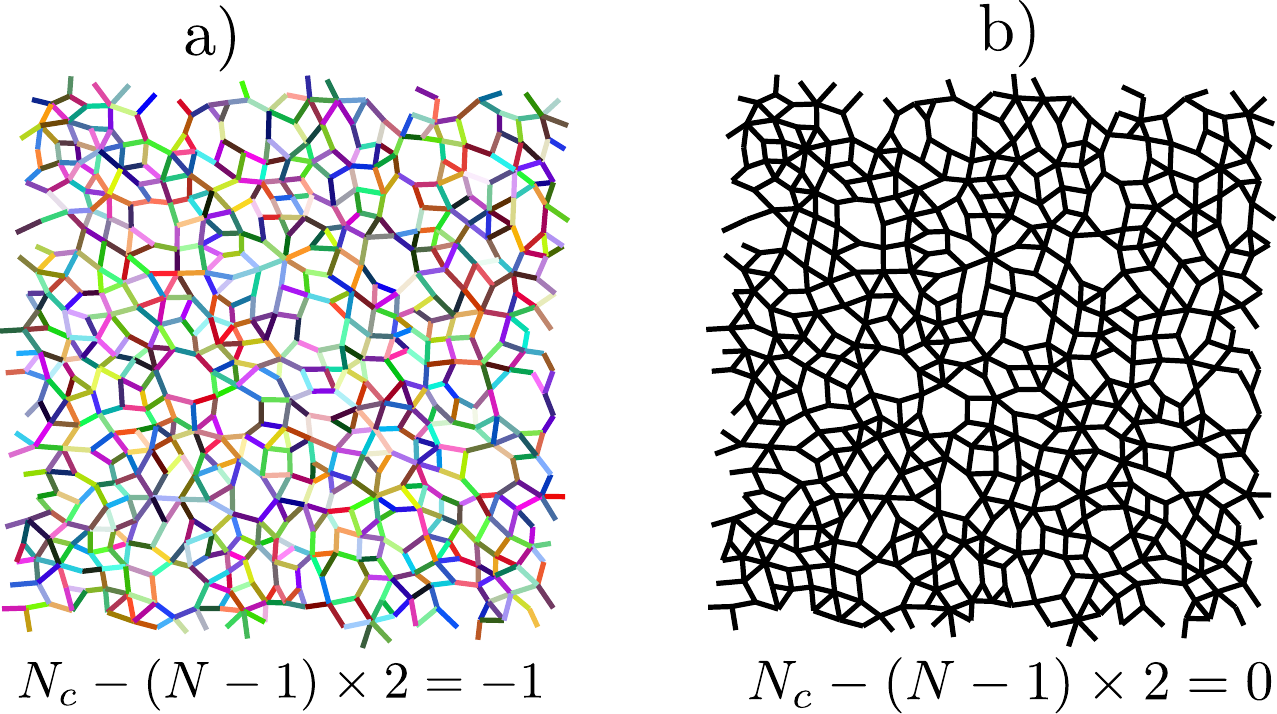}
    \caption{\textbf{Rigidity transition in sheared frictionless disk packings.} Pebble game analysis 
    on the isostatic networks yields a single rigid cluster consisting of the whole system (\textbf{b)}. Removal of 
    one bond from that network results in a complete loss of rigidity, with the pebble game decomposing the system into 
    multiple small rigid clusters indicated by the different colors (\textbf{a)}).} 
    \label{fig:fricitonless_rigidity}
\end{figure}
\begin{figure}
    \centering
    \includegraphics[scale=0.38]{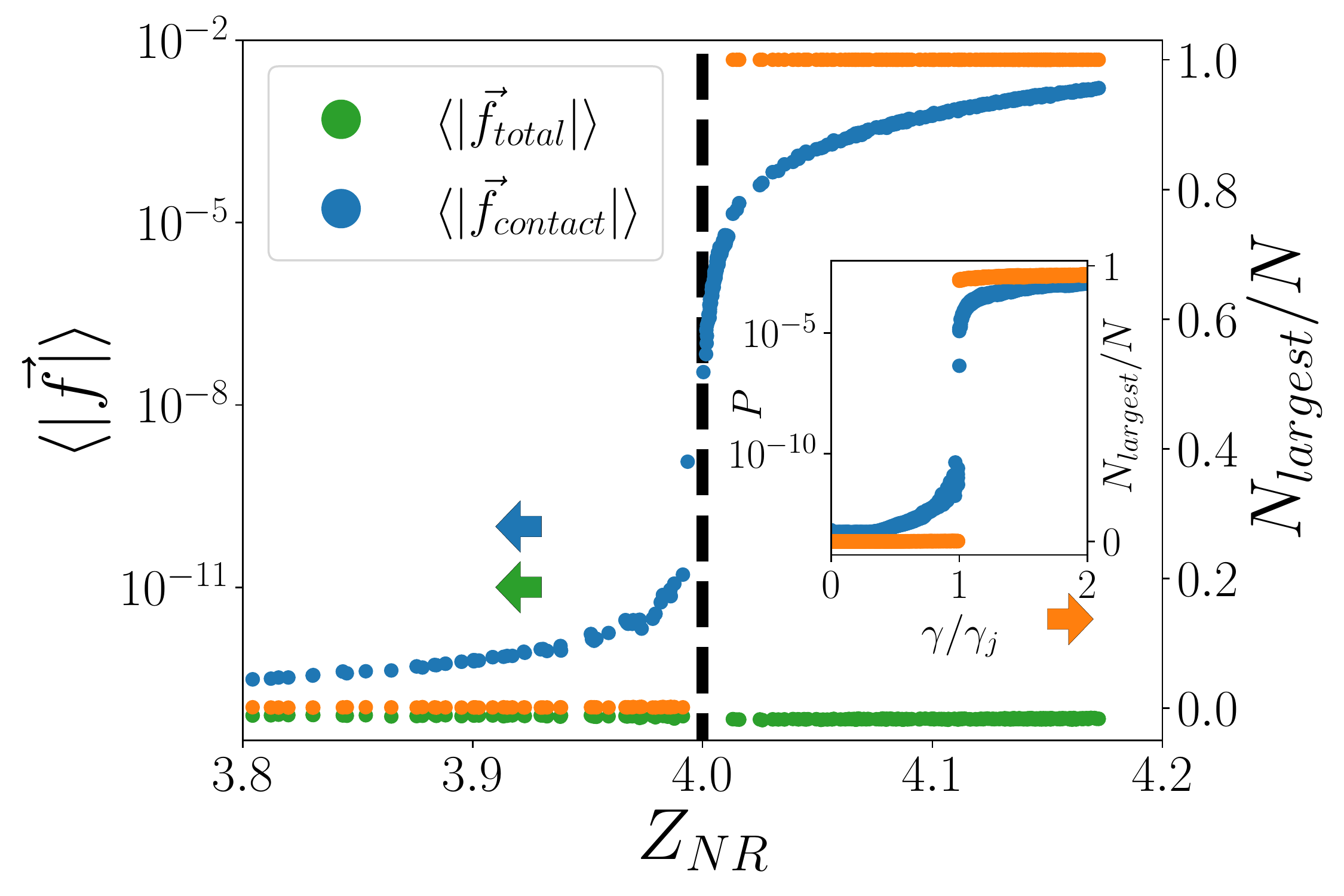}
    \caption{\textbf{Rigidity transition associated with shear jamming in frictionless 
    systems.}  Rigidity transition can be seen as a discontinuous jump in the size of the 
    largest cluster. Inset shows pressure $P$ vs strain $\gamma$ and the rigidity 
    transition. The transition occurs at the isostatic value of the non-rattler contact number, $Z_{NR} = 4$.}
    \label{fig:frictionless_shearjamming}
\end{figure}

First, we discuss the results of the frictionless system, for which above jamming, energy minimization cannot remove all the overlaps in the system, resulting in finite forces. As discussed in \cite{babu2021dilatancy,babu2022criticality}, configurations are iso-static ($N_c = (N-1)\times2$ after removing rattlers) at the jamming point. $(k=2,l=2)$ pebble game analysis of isostatic configurations shows that 
the whole system is made up of a single rigid cluster, as shown in Fig. \ref{fig:fricitonless_rigidity} (a). Removing a single bond from this system leads to loss of rigidity, as shown in  Fig. \ref{fig:fricitonless_rigidity} (b).  The results of this analysis are summarized in Fig. 
\ref{fig:frictionless_shearjamming}.  The shear jamming transition can be identified by 
the presence of finite contact forces as well as by $Z_{NR}$. The rigidity transition occurs at the jamming transition point  and is characterized by a discontinuous jump in 
the size of the largest cluster. This strongly discontinuous rigidity transition is therefore common for frictionless isotropic and shear jamming. 

 

Next, we discuss the results from finite rate shear of frictional systems for  
shear rates $\dot{\gamma}=10^{-6},10^{-7},10^{-8},10^{-9},10^{-10}$. The 
main observation from this set of simulations is that the rigidity transition associated 
with shear jamming becomes ``sharper''  as one reduces the 
shear rate, an observation also made in \cite{henkes2016rigid}. 
\begin{figure}
    \centering
    \includegraphics[scale=0.255]{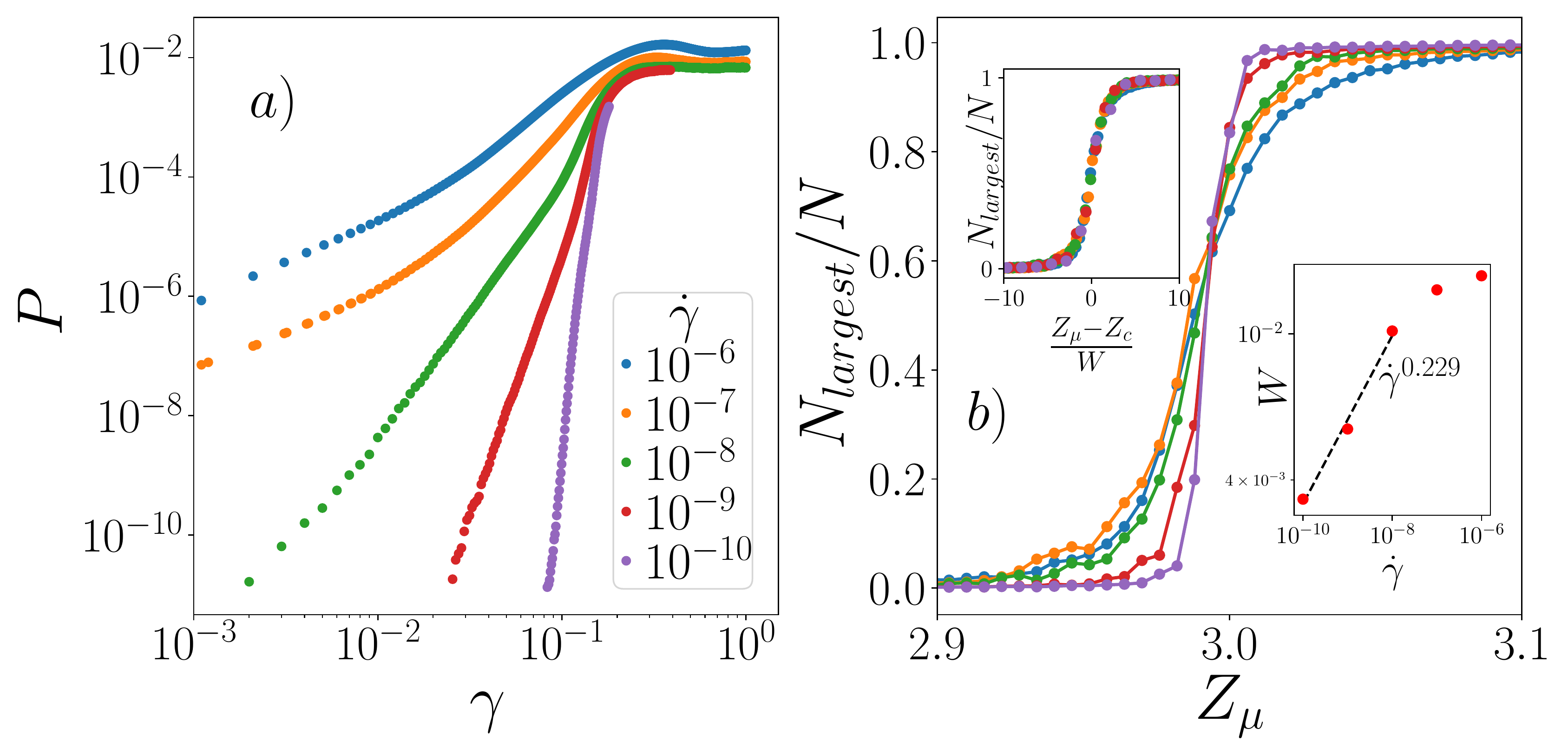}
    \caption{\textbf{Finite rate shear for $N=16384$ with $\mu=1$.} \textbf{a)} Pressure $P$ vs $\gamma$. \textbf{b)} Fraction of 
    the largest rigid cluster with the total number of particles as a function of $Z_{\mu}$. As $\dot{\gamma}$ is reduced the transition becomes ``sharper''. \textbf{Inset top left:} Data from different shear rates collapse onto each other when scaled by the ``width'' $W$ of the transition region. \textbf{Inset lower right:} The width of 
    the transition region obtained by fitting the data. Dependence of $W$ for the three smaller shear rates on $\dot{\gamma}$ can be described using a power-law suggesting that the transition becomes discontinuous as $\dot{\gamma} \rightarrow 0$. }
    \label{fig:gdot_dependenence}
\end{figure}
As shown in Fig. \ref{fig:gdot_dependenence} (a), the  increase in pressure $P$ with strain 
is noticeably sharper for smaller shear rates. To characterize the rigidity of these configurations we follow \cite{henkes2016rigid,liu2021spongelike,vinutha2019force} and use the $(k=3,l=2)$ pebble game on the contact network. Note that in the finite rate simulations, we do not simulate the system till it 
achieves force balance, and therefore for jammed as well as unjammed configurations, the net forces on the disks are finite.
We use a threshold $\delta$ to identify mobilized contacts - if $\frac{|\vec{f}_t|}{|\vec{f}_n|} > \mu - \delta $ 
then the contact is mobilized. For simulations with $\mu=1$, very few of our contacts are sliding  and the choice of $\delta$ does not significantly affect the results presented. The choose 
 $\delta = 10^{-12}$ for the results in the main text. A discussion on the choice of $\delta$ is included in the SM section VI.
\begin{figure}
    \centering
    \includegraphics[scale=0.25]{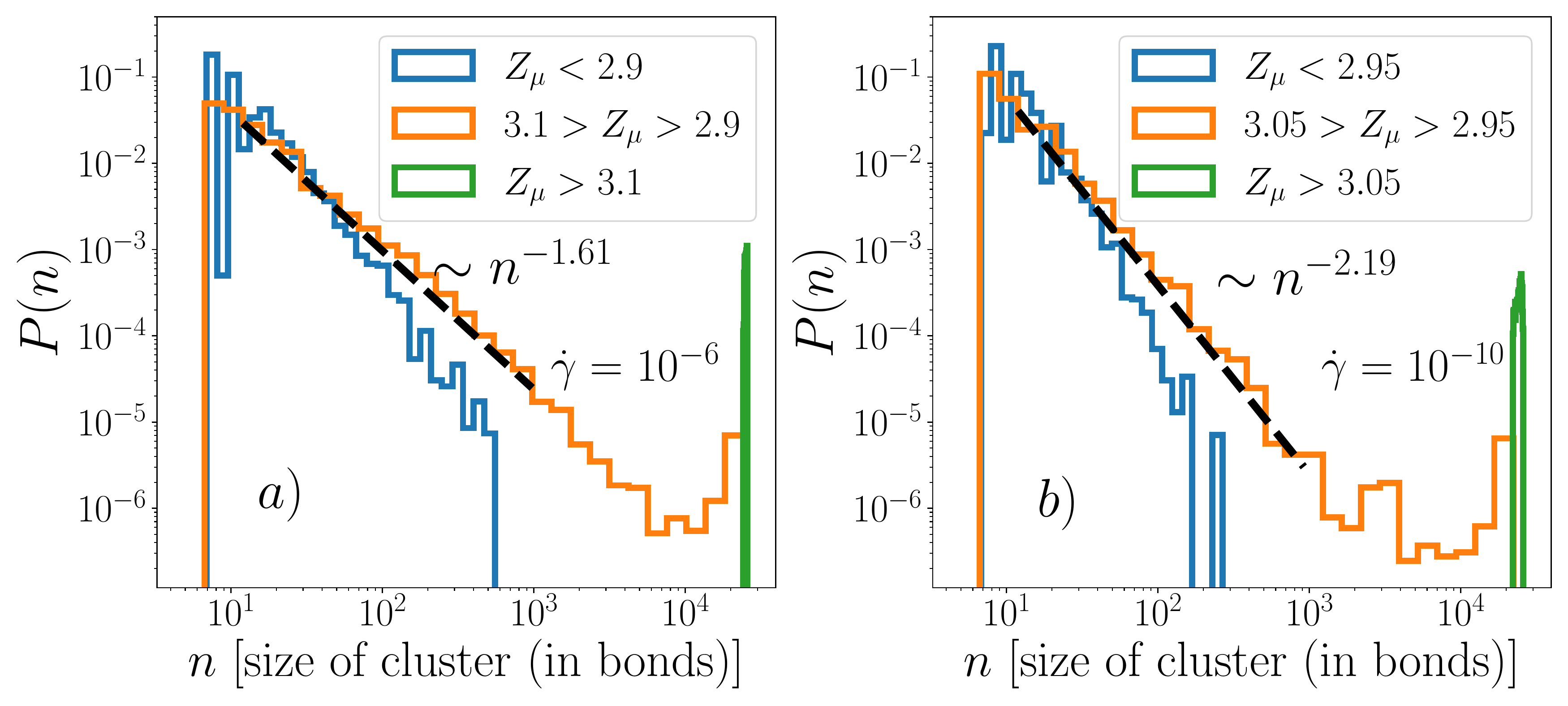}
    \caption{\textbf{Comparison of cluster size distribution between high and low $\dot{\gamma}$ studied.} \textbf{a)} $\dot{\gamma}=10^{-6}$ and \textbf{b)} $\dot{\gamma}=10^{-10}$. 
    Comparing the distribution 
    of cluster sizes for the range covering $3$, we see that $\dot{\gamma}=10^{-6}$ shows a broader 
    distribution  {compared to the one at $\dot{\gamma}=10^{-10}$ as quantified by the exponent 
    characterizing the power law distribution}, indicating that the transition becomes discontinuous as the shear rate vanishes. The distribution corresponding to a given region in $Z_{\mu}$ is calculated by considering the sizes of 
    all rigid clusters in a configuration with $Z_{\mu}$ in that region.}
    \label{fig:cluster_size_distribution_finite_rate}
\end{figure}
Even though the system is not in force balance when sheared at a finite rate, we identify rattlers as 
particles with just one contact and remove them recursively.  For the remaining contact network, 
we perform pebble game analysis and show in Fig. \ref{fig:gdot_dependenence} (b)  the size of the largest rigid cluster as a function of the average contact number $Z_{\mu}=Z_{NR}-n_m$. The transition becomes sharper as one reduces $\dot{\gamma}$, and interestingly, the transition 
occurs at $Z_{\mu} \approx 3$, the isostatic value, for all shear rates. We fit the data 
using the logistic function $f(x)=\left[1+e^{-\frac{x-Z_c}{W}}\right]^{-1}$ (as a reasonable but arbitrary choice) and use $W$ as a 
measure of the width of the transition region. As the top left inset in \ref{fig:gdot_dependenence} (b) shows, the data can be collapsed using the fit values, with $Z_c\approx 2.99$. In the lower right inset, we show the behavior of $W$, whose dependence on $\dot{\gamma}$ can be described by a power law that implies that the transition becomes discontinuous at $\dot{\gamma}\rightarrow 0.$ To our knowledge, this has not been reported for  shear jamming transition. 


Next, we study the cluster size distribution as shown in Fig. \ref{fig:cluster_size_distribution_finite_rate} 
for the largest and the smallest $\dot{\gamma}$ 
studied. For both cases, we divide the region studied  {(in $Z_{\mu}$)} into three regimes -- before the jamming transition, 
 a regime covering the transition, and after the transition -- and compute the distribution of the rigid cluster sizes separately for each of them. 
 The distributions in the regime covering the transition are quantified by an exponent characterizing the power-law distribution of the rigid clusters. For $\dot{\gamma}=10^{-6}$, the exponent is $-1.61$ and for $\dot{\gamma}=10^{-10}$, the exponent is $2.19$. While the transition in this regard appears continuous for both the shear rates studied, the distributions become progressively narrower as the shear rate decreases. The corresponding curves for the frictionless and frictional quasistatic shear show a faster than power law decay below the rigidity transition.  
 We also calculate $P_\infty$, the probability that a given disk belongs to a system spanning (percolating) rigid cluster, which is shown in the SM section IV. The  $P_\infty$ curves become progressively step-like with decreasing shear rate. Thus, we conclude that the appearance of a continuous transition is associated with the finite shear rates and absence of force/torque balance, rather than being an indication of the intrinsic nature of the shear jamming transition, or the presence of friction.

\begin{figure}
    \centering
   \includegraphics[scale=0.31]{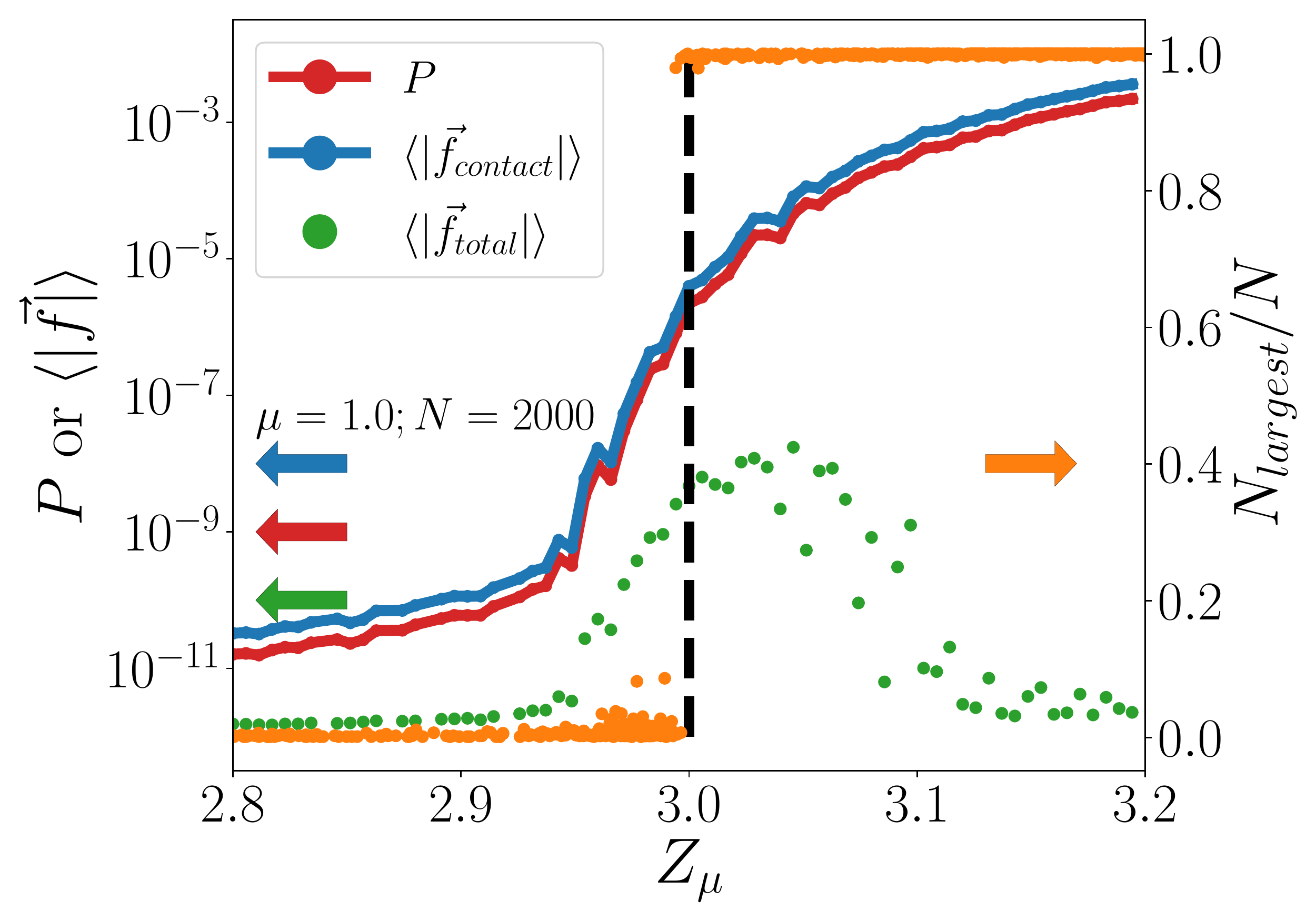}
    \caption{\textbf{Rigidity analysis of quasi-statically shear jammed frictional disks}. The 
    size of the largest rigid cluster discontinuously jumps to equal the system size 
    as $Z_{\mu}$ crosses $3$, the iso-static value. The contact forces and pressure show a more gradual change, but the behavior of the net force on the disks reveals this to be a result of incomplete convergence, as indicated by the average of the net force of individual disks.  
    }
    \label{fig:jamming_forces_rigid_cluster}
\end{figure}

To underscore our conclusions, we next consider quasistatic shearing of frictional disks,  
which is performed by applying an affine 
transformation and relaxing the system using DEM till the system reaches force balance. 
As noted before, the relaxation near the jamming transition is very slow and therefore 
it is hard to generate force-balanced configurations near the jamming transition \cite{vinutha2020timescale,shivers2019scaling}. 
Given configurations that are fully relaxed,  we define rattlers as particles that do not have finite forces acting on them. Disks with a single contact cannot sustain a non-zero force on that contact, which we remove  
recursively. In addition, given a friction co-efficient $\mu$, disks with two contacts can be in force balance with finite forces only if the angle $\theta$ between the two contacts is large enough. 
If $\mu<\tan(\frac{\pi}{2}-\frac{\theta}{2})$, these contacts cannot carry forces  {(see SM section III)}, and are this also removed recursively. 

These configurations are analyzed using the $(k=3,l=2)$ pebble game, and the results are shown in Fig. 
\ref{fig:jamming_forces_rigid_cluster}. As $Z_{\mu}$ crosses the isostatic value $3$, the largest rigid cluster encompasses 
the whole system, exhibiting a striking similarity 
with the behavior found for the frictionless case (Fig. \ref{fig:frictionless_shearjamming}). This observation is even more 
remarkable when one considers the behavior of the contact forces or pressure, {\it vs.} $Z_{\mu}$, which show a more rounded 
change, as a result of the difficulty of converging to force balanced configurations, as indicated by the non-monotonic 
behavior of the net forces acting on the disks. $P$, $\langle |\Vec{f}_{contact}| \rangle$ and $\langle |\Vec{f}_{total}| 
\rangle$ shown are average values computed from all configurations having a given value of $Z_{\mu}$. $N_{largest}/N$ is a scatter plot from all trajectories.




Before closing, we briefly compare our results and conclusions with previous work mentioned earlier. 
While the conclusion in \cite{henkes2016rigid} differ from ours, the 
sharpening of the rigidity transition has also been noted in 
\cite{henkes2016rigid}. In \cite{vinutha2019force}, shear was applied to 
frictionless disk assemblies before friction was included in the rigidity 
analysis. While this procedure captures many features of sheared frictional 
disks, like the  anisotropy and the emergence of a contact network that 
supports jamming in the presence of friction, subtle but important 
differences in the organization of contacts exist. Specifically, using the 
procedure of \cite{vinutha2019force}, the fraction of redundant bonds rises 
continuously from below the isostatic contact number, as shown in the SM 
Section VII, whereas they are strictly zero below the frictional jamming 
point. The absence of redundant bonds before the rigidity transition is a 
characteristic feature of jamming, as compared to rigidity percolation in 
spring networks and other systems \cite{chubynsky2006self}. Our results 
differ from the analysis of experimentally sheared disk packings in 
\cite{liu2021spongelike}, for which we do not have a ready explanation, since 
the experimental protocol should be expected to closely agree with the 
quasistatic shear we employ, an inconsistency that needs to be further 
investigated. 

In summary, our results unambiguously demonstrate that the rigidity transition associated with shear jamming in both frictionless and frictional disk packings is discontinuous in nature, when conditions of force and torque balance are met. Thus, the nature of the emergence of rigidity is the same for isotropic and shear jamming. Features that suggest a continuous transition are associated with partial relaxation of unbalanced forces, as our results for finite shear rate demonstrate, but such behavior approaches discontinuous change as the shear rate vanishes. Our results thus establish a key additional element in the shared phenomenology of isotropic and shear jamming. 

We thank Sumantra Sarkar, Sanat Kumar, Karen Daniels and Silke Henkes for useful discussions. We acknowledge support from the Thematic Unit of Excellence on Computational Materials Science (TUE-CMS) and the National Supercomputing Mission facility (Param Yukti) at the Jawaharlal Nehru Centre for Advanced Scientific Research (JNCASR) for computational resources. D.B. acknowledges support from the National Science Foundation (grant no. DMR-2046683) and the Alfred P. Sloan Foundation. S.S. acknowledges support through the JC Bose Fellowship (Grant No. JBR/2020/000015) from the Science and Engineering Research Board, Department of Science and Technology, India. 

\end{document}